\documentclass[
]{ceurart}

\usepackage{listings}
\begin{document}

\copyrightyear{2024}
\copyrightclause{Copyright for this paper by its authors.
  Use permitted under Creative Commons License Attribution 4.0
  International (CC BY 4.0).}

\conference{MMSR2024 Workshop}

\title{Smart Multi-Modal Search: Contextual Sparse and Dense Embedding Integration in Adobe Express}



\author{Cherag Aroraa}[
email = charora@adobe.com
]

\author{Tracy Holloway King}[
email = tking@adobe.com]

\author{Jayant Kumar}[
email = jaykumar@adobe.com
]

\author{Yi Lu}[
email = yil@adobe.com
]

\author{Sanat Sharma}[%
email=sanatsha@adobe.com
]

\author{Arvind Srikantan}[
email = asrikantan@adobe.com
]

\author{David Uvalle}[
email = duvallezeped@adobe.com
]

\author{Josep Valls-Vargas}[
email = jvallsvargas@adobe.com
]

\author{Harsha Vardhan}[
email = hmatadaallam@adobe.com
]

\address{Adobe Inc.}

\begin{abstract}
As user content and queries become increasingly multi-modal, the need for effective multi-modal search systems has grown. Traditional search systems often rely on textual and metadata annotations for indexed images, while multi-modal embeddings like CLIP enable direct search using text and image embeddings. However, embedding-based approaches face challenges in integrating contextual features such as user locale and recency. Building a scalable multi-modal search system requires fine-tuning several components. This paper presents a  multi-modal search architecture and a series of AB tests that optimize embeddings and multi-modal technologies in Adobe Express template search. We address considerations such as embedding model selection, the roles of embeddings in matching and ranking, and the balance between dense and sparse embeddings. Our iterative approach demonstrates how utilizing sparse, dense, and contextual features enhances short and long query search, significantly reduces null rates (over 70\%), and increases click-through rates (CTR). Our findings provide insights into developing robust multi-modal search systems, thereby enhancing relevance for complex queries.

\end{abstract}


\maketitle

\section{Introduction}

\begin{figure}[htb]
  \centering
  \includegraphics[width=3in]{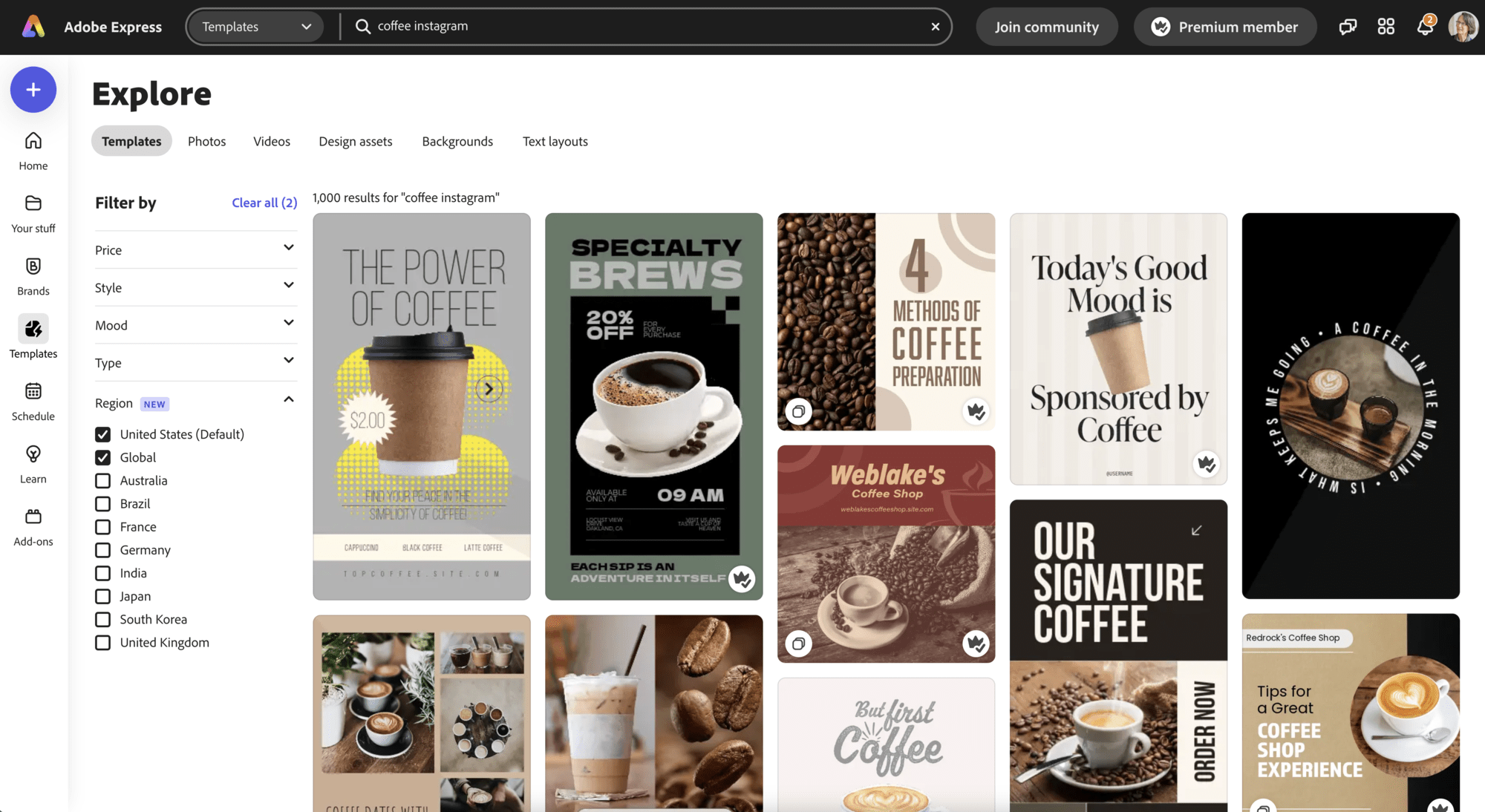} 
  \hspace*{1em}
  \includegraphics[width=3in]{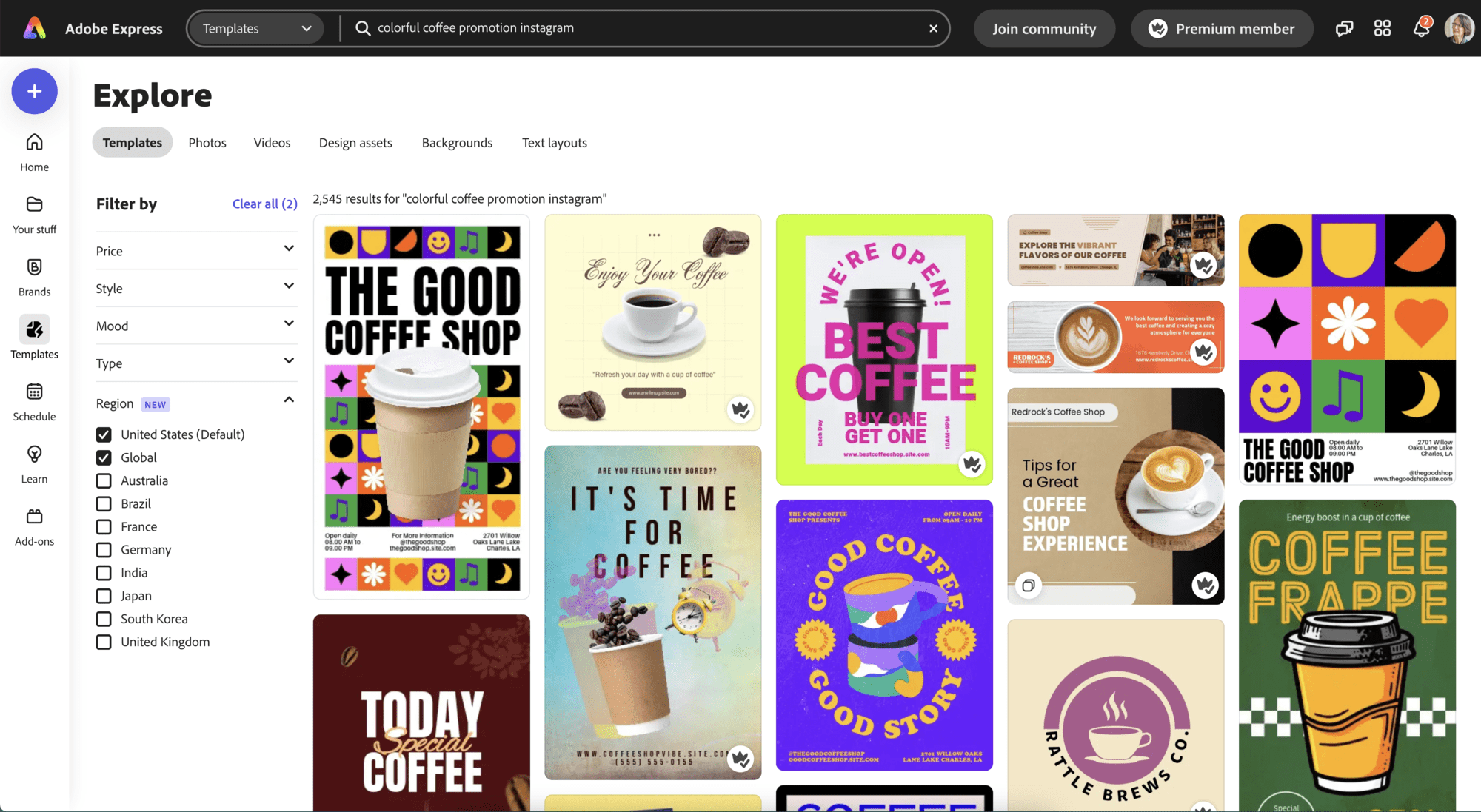}
  \caption{Examples of Express template search.  Left: head query \textit{coffee instagram}. Right: tail query \textit{colorful coffee promotion instagram}}
  \label{fig:express-ex}
\end{figure}

 For search over images and multi-modal content, industry search systems traditionally rely on textual and metadata annotations added to indexed images. However, multi-modal embeddings like CLIP \citep{clip2021} enable direct search of image content using text and image embeddings, allowing for direct text-to-image and image-to-image search. While pure embedding-based approaches facilitate  content understanding, they struggle with integrating contextual features like user locale and recency into retrieved results. Building a production-grade, scalable, multi-modal search system involves carefully tuning several components. This paper describes a series of AB tests conducted to leverage embeddings and other multi-modal technologies in search for Adobe Express templates. These templates are complex multi-modal (and multi-page) documents, containing images, text and rich metadata (section \ref{sec:modelsdata}). Figure \ref{fig:express-ex} shows the Adobe Express template search for a head query and a tail query, where the templates are displayed to the user as images and template metadata drives the left rail filters.

To improve text search for templates, integrating embeddings required decisions as to:

\begin{itemize}
    \item Which embedding model(s) to use
    \item Whether to leverage embeddings for matching (recall), ranking, or reranking
    \item Whether to use dense or sparse embeddings
    \item Whether head and tail queries should be treated identically
    \item Whether embeddings should be used for null and low recovery or everywhere
\end{itemize}

\noindent Other than which embeddings to use, these decisions were driven by latency concerns and by constraints on integration with Elasticsearch, which was the existing inverted index used for Express template search.  With an ever-increasing collection of $\sim$300,000 templates, dense embeddings could not be used for matching due to the number of scoring calculations, which leads to high latency. This restricted dense embeddings to (re)ranking, where only a small ($<$10K) number of top templates had to be scored, and to scenarios like null and low recovery and long tail queries, where the additional latency was worth the improved relevance. In addition, certain types of queries performed better with keyword search, especially those around design type (e.g.\ poster, Instagram reel) and format (e.g.\ still, animated, video).

To determine the optimal combination, we took an iterative approach with a series of evaluations and AB tests.  We started with existing models with single integrations and then built on these to improve remaining relevance issues.  This paper first overviews the data and models used (section \ref{sec:modelsdata}) and then discusses the experiments and how the decisions were made for each of these (section \ref{sec:experiments}).

\section{Models and Data}
\label{sec:modelsdata}

This section describes key data and models used for the Express template recall and ranking. The templates themselves contain rich image, text, and metadata. The standard search behavioral data (e.g.\ impressions, clicks) are available, as well as certain application-specific behavioral data (e.g.\ number of edits, number of exports). These are briefly described in section \ref{sec:templatedata}. In addition, we have two types of multi-modal models: two CLIP text-image models (section \ref{sec:clipAdobeCLIP}) and an intent-based model (section \ref{sec:mmCKG}).

\subsection{Template Data}
\label{sec:templatedata}

\begin{figure}[htb]
\begin{tabular}{ll}
\raisebox{-.5in}{\includegraphics[height=1.5in]{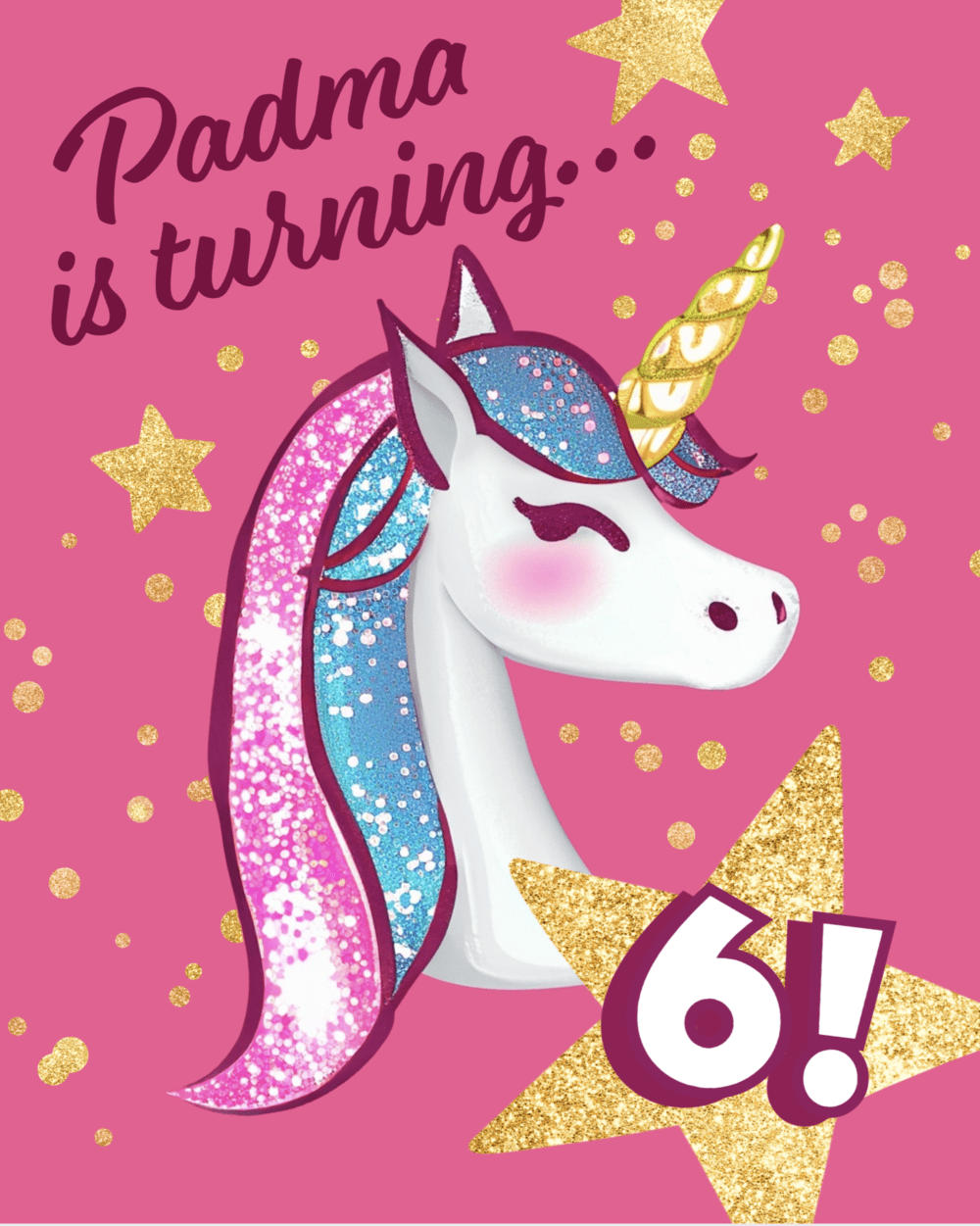}} & \begin{tabular}{l}
Title: Pink Unicorn Birthday Party Instagram Portrait Post\\
Topics: confetti, fantasy, glitter, gold, kids, sparkle, star, unicorn\\
Mood: happy, joyful; Style: bright\\
Region: all; Language: en-US; Date: 2023-12-12; Behavior: still; License: premium\\
AI-inferred: AdobeCLIP embedding, Multi-modal CKG embedding,\\
\hspace*{1em}CKG symbolic intent, autotags\\
Behavioral: Search impressions, clicks, edits, exports\\
\end{tabular}
\end{tabular}
\caption{Sample Express template data available for search matching and ranking}
\label{fig:templatedata}
\end{figure}

Express templates are rich objects which contain many visual layers and text boxes. These can also be viewed as images, e.g.\ those that are displayed in search (Figure \ref{fig:express-ex}). In addition, templates have titles provided by the template designers as well as filter information such as design type, style, mood, region, and price (free/premium). Additional information is inferred about each template including multi-modal embeddings, user intents, and image tags. Finally, aggregated behavioral data such as impressions, clicks, edits (number of edits users make to the template in order to personalize them) and exports (number of times the template is exported after editing) are available. An example template with its data is shown in Figure \ref{fig:templatedata}.

\subsection{Image-Text CLIP Embeddings}
\label{sec:clipAdobeCLIP}
CLIP \citep{clip2021} embeds images and text in the same space.  This allows for embedding-based search of images using text queries. There are several off-the-shelf CLIP models available.  However, for Express template search and for other visual asset search like Adobe Stock, we needed a model that: (1) worked on short text (queries) as well as long text (captions); (2) covered five languages (English, French, German, Japanese, Korean); (3) performed well on high-quality image data for templates, photographs and illustrations; (4) had a sparse version as well as the dense vectors. To meet these requirements, we trained a CLIP-architecture model on  Adobe-licensed image-text data. The text model was particularly important since the training focused on Adobe vocabulary, shorter text, and multiple languages; the training architecture was inspired by \cite{carlsson-etal-2022-cross}.

There are many ways to improve the latency when using embeddings with large numbers of assets. However, the approximate  methods reduce accuracy because the list of assets whose embeddings are closest to the query embedding is not exact. Once a smaller set of  embeddings is selected (e.g.\ by using the top \textit{n} embeddings from the approximate scoring), then the dense embedding can be used to get more accurate scores for a final ranking.
We used a sparsification method which allows the embeddings to be used similar to keywords in the existing index.\footnote{Other approaches are described in \cite{googlesparse,pqcodes,kusupati2024matryoshka}.}\  An example of this is shown in Table \ref{tab:rae}.

\begin{table}[htb]
\caption{Dense and Sparse representations of embeddings with sample scoring for sparse embeddings. Dense embeddings are shown with 2048 dimensions. Sparse embeddings have more dimensions (here 8192) but most of the dimensions have no values.}
\begin{tabular}{|l|ll|l|l|lll|}
\hline
\multicolumn{3}{|c|}{Dense Embedding Example} & \multicolumn{5}{|c|}{Sparse Embedding Example}\\\hline
Dim. & Img.\ 1 & Img.\ 2 & Dim. & Query =  & Img.\ 2 & Img.\ 3 & Img.\ 4\\
& & & & Img.\ 1 & & &\\\hline
1 & 0.11 & 1.23 & 1 & --- & 1.12 & --- & ---\\
2 & 1.21 & 0.42 & 2 & --- &--- &  0.81 & ---\\
 3 & 0.15 & 0.53 & 3 & 1.16 & 0.83 & --- & 0.64\\
4 & 0.22 & 2.25 & 4 & --- & --- & 1.83 & ---\\
$\ldots$ && & $\ldots$ &  &  &  & \\
2048 & 2.17 & 0.64& 8192 & 0.13 &---  & --- & 0.01\\
\hline
\end{tabular}

\vspace*{1em}
\begin{tabular}{|ll|}
\hline
\multicolumn{2}{|l|}{Sparse Embedding Matching and Scoring}\\\hline
\multicolumn{2}{|l|}{Matches for query image 1:}\\
&image 2: 1 dimension (dimension 3)\\
&image 3: 0 dimensions\\
&image 4: 2 dimensions (dimension 3 \& 192)\\
\multicolumn{2}{|l|}{Scoring for ranking:}\\
& image 2: 1.16 * 0.83 = 0.96\\
& image 4: 1.16 * 0.64 + 0.13 * 0.01 = 0.74\\
\hline
\end{tabular}

\label{tab:rae}
\end{table}

The dense embeddings have values for every dimension (2048 dimensions in table \ref{tab:rae}). The sparse version, which is derived from the dense one, has more dimensions (8192 in table \ref{tab:rae}) but most of them have no values.  For a  query,  only assets which match at least \textit{n} dimensions are returned.  In the example, \textit{n} is set to 1 and so image 2 matches dimension 3 and image 4 matches dimensions 3 and 8192, while image 3 is not matched.  The scoring is the sum of the matched dimension values weighted by the score of that dimension for the query.  This sparse encoding for matching and scoring is extremely fast, but comes at the cost of lower accuracy. 

The Adobe-specific model (AdobeCLIP) was evaluated for Adobe Express and Stock content, baselining against the off-the-shelf CLIP versions. For large scale evaluation, a stratified sample of held-out search queries were used for semantic search against an index of CLIP and AdobeCLIP embeddings for Express and Stock content. Past clicked assets were considered relevant, non-clicked items irrelevant.  In addition, we selected titles (generally 5--15 words) for Express templates and Stock images and used them as long queries, measuring the position of the asset which originally had that title. These  methods provide a lower bound on performance since many of the non-clicked items are relevant and titles often matched multiple high-ranking images.  These two approaches allowed us to quickly compare different versions of AdobeCLIP against one another and against CLIP.  Once the AdobeCLIP model outperformed CLIP and earlier AdobeCLIP versions, we  manually inspected results for a subset of held-out queries.

\subsection{Multi-Modal Creative Knowledge Graph}
\label{sec:mmCKG}

In addition to learning representations of the content via AdobeCLIP, we found mapping the content's intent to discrete nodes improved recall and explainability and  allowed for downstream-recommendation tasks, similar to \cite{le2024combiningembeddingbasedsemanticbasedmodels}. However, we  discovered that self-supervised models like AdobeCLIP, which were trained on asset-caption and asset-query data like Adobe Stock and Adobe Express failed to accurately map the asset's intent to short discrete labels. To accomplish, this we created a ``Creative'' Knowledge Graph (CKG) \citep{US11645095B2,US11775734B2,2023font} containing over 100K nodes focusing on Adobe-specific user intents.  We then trained a multi-modal transformer (MM-CKG) specializing in mapping assets to these discrete nodes using supervised contrastive training.
We mined concepts for events, actions, objects, moods, canvas types, colors, and backgrounds  to get a robust understanding of an asset's content. For example, \textit{actions} has subtypes of \textit{run, dance, $\ldots$}; \textit{events} has subtypes of \textit{birthday, graduation, wedding, seasonal, $\ldots$}; in turn  \textit{events$\mid$seasonal} has subtypes of \textit{Halloween, Thanksgiving, 4th of July, $\ldots$}.


To train the model, we created sequence-wise self-attention blocks inspired by \cite{liu2021cmaclipcrossmodalityattentionclip}. We built our model on top of a base CLIP backbone and then added a sequence-wise attention block on top that takes in the hidden states from the last layer of the CLIP backbone that runs through a couple layers of multi-headed transformer blocks. 
We utilized the $T_{cls}$  and $I_{cls}$ outputs from the sequence-wise attention heads as the final representation of the input image and text modalities.
\begin{figure}[htb]
 \centering
 \includegraphics[width=3.25in]{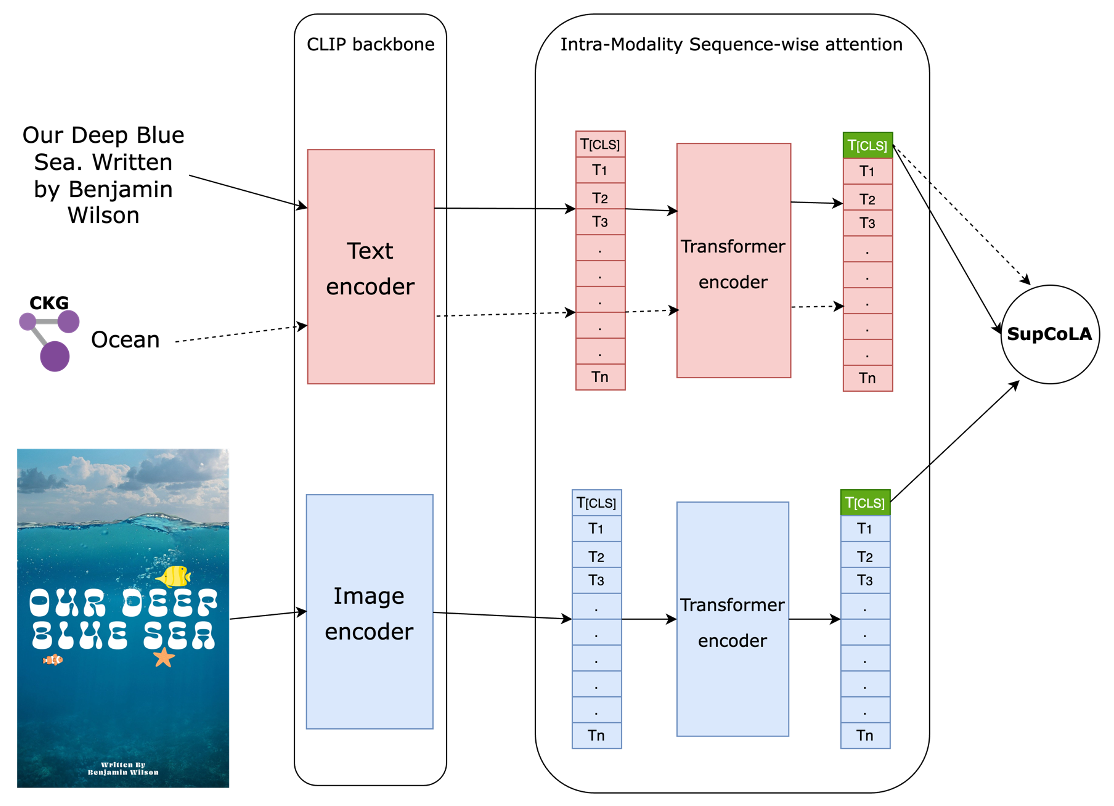}
 \caption{ MM-CKG  uses supervised contrastive learning with SupCoLA loss  for label alignment. This allows the model to bring short labels focusing on the overall intent of the asset closer to the content embeddings.}
 \label{fig:mmckg}
\end{figure}

\subsubsection{Supervised Contrastive Loss (SupCoLA)}

We devised our loss function with the following requirements:

\begin{enumerate}
    \item Alignment to labels: Ensure that the image and text in the training process were  close to the label embeddings. 
    \item Ability to handle multiple positives in a batch: Traditional contrastive learning (InfoNCE loss \cite{oord2019representationlearningcontrastivepredictive}) assumes that for a given pair in a batch, all other pairs are negatives. However, when learning alignment with labels, multiple rows with the same label may be present in a batch. The loss function should not penalize these rows during loss computation.
    \item Ability to have multiple labels per row: Some rows  have multiple labels. For example, for the prompt, \textit{boy is sitting on a beach with his dad for father’s day}, there are multiple concepts: the creative intent \textit{father’s day}, the scene objects \textit{boy} and \textit{beach}, and the background \textit{beach background}.
\end{enumerate}

\noindent Our resulting loss function, Label-Aligned Supervised Contrastive Loss, is based on SupCon loss \citep{khosla2021supervisedcontrastivelearning} where we pass image, text and label embeddings as anchor features as well as contrast features. 

\begin{equation}
\mathcal{L}^{\text{sup}} = \sum_{i \in I} \mathcal{L}_{i}^{\text{sup}} = \sum_{i \in I} \left( -\frac{1}{P(i)} \sum_{p \in P(i)} \left[ \sum_{v \in j(p)} \log \frac{\exp (\mathbf{z}_{i} \cdot \mathbf{z}_{v} / \tau)}{\sum_{n \in A(i)} \exp (\mathbf{z}_{i} \cdot \mathbf{z}_{n} / \tau)} \right] \right)
\end{equation}

\noindent where $I$ is the mini batch, $i$ is the index of anchor sample in the batch, $A(i)\equiv I\setminus{i}$ is the set of all samples $n$ in the batch that have distinct index than the anchor $i$, $j(p)$ is the set of all positives $p$ in the batch that have the same label as anchor $i$ and are views of $p$. Views of sample $p$ denote the embeddings for the label, image and text modalities.


Why do we have two domain-specific multi-modal embeddings (AdobeCLIP and MM-CKG)? These target and excel at  different use cases, both of which are important for template search relevance.  MM-CKG is better at determining the underlying key intent from a query and at specific scene object detection.  AdobeCLIP is better at color and layout understanding.

\section{Iterative Experiments}
\label{sec:experiments}

This section describes the series of on-line experiments conducted to improve the Express template multi-modal search.  We focus primarily on experiments involving multi-modal embeddings, but include one  experiment that leveraged multi-modal content under the hood but text at query time.

\begin{figure}[htb]
  \centering
  \includegraphics[width=4in]{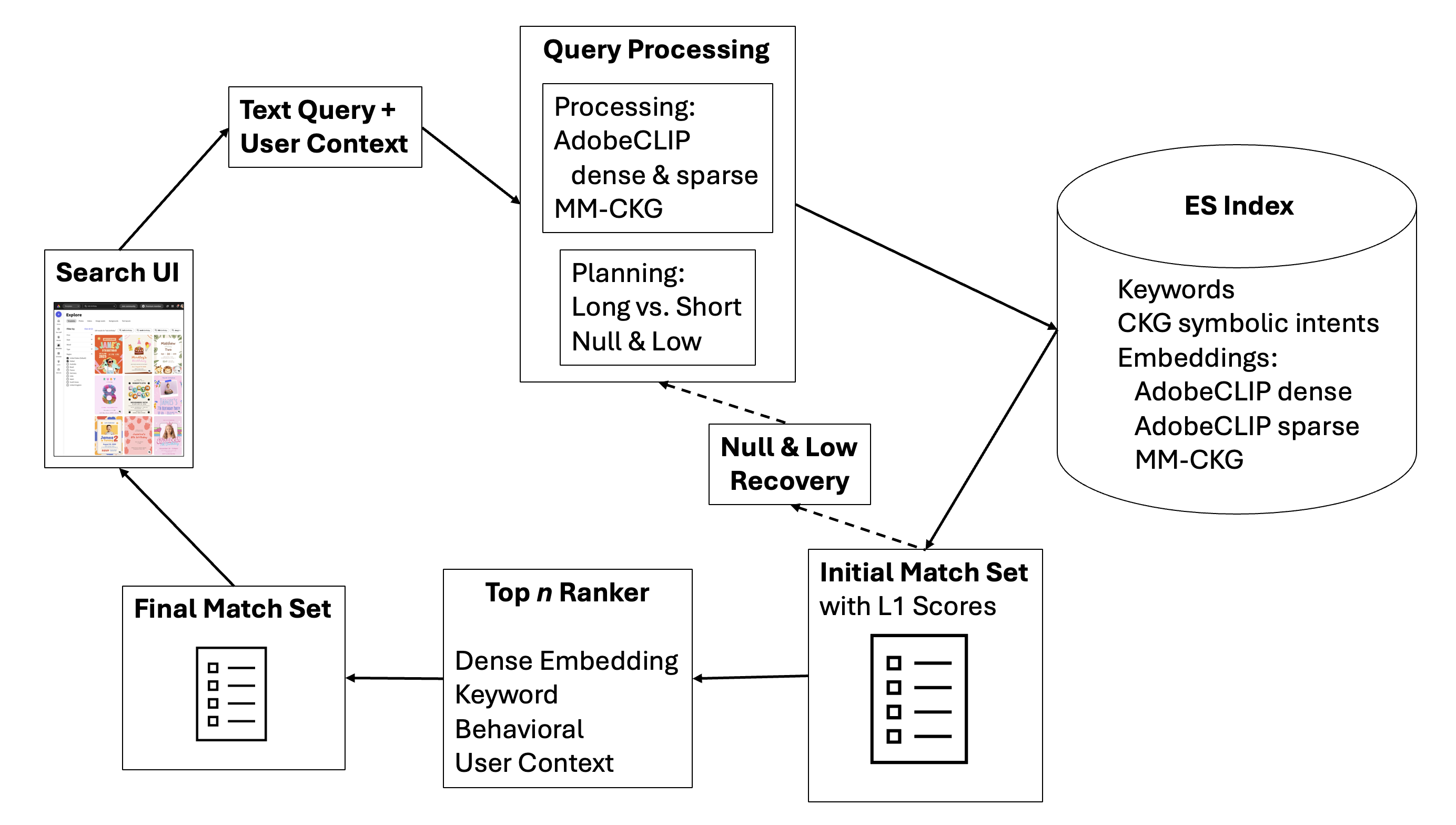} 
  \caption{High-level search architecture for Express template multi-modal search}
  \label{fig:searcharchi}
\end{figure}

The Express template search uses a standard architecture for relevance (Figure \ref{fig:searcharchi}). It is built on Elasticsearch. There is an initial matching (recall) step to retrieve documents which broadly match the user's query. This  step  uses keyword-style matching against text and metadata and includes an initial low latency scoring. Matching using sparse AdobeCLIP embeddings for all queries (section \ref{sec:exp:rae}) and dense multi-modal CKG embeddings for long queries (section \ref{sec:exp:mmckg}) were also added.  If not enough results are found, null and low recovery occurs, including a speller (not discussed in this paper; see \citep{speller2023}) and the use of symbolic CKG intents (section \ref{sec:exp:ckgnulllow}). The top 10K templates from the initial match set are then reranked using a much broader set of features. This includes dense multi-modal embeddings as well as the usual discrete features such as BM25, locale, language, and aggregated behavioral data. 

To improve relevance, and hence search click-through rate and export rate, we took an iterative approach to learn how to optimally leverage our multi-modal understanding, especially the multi-modal embeddings. Each of these are discussed in detail below. Since we were redesigning the search system, including the retrieval and ranking platform, some of the experiences were evaluated extensively offline and then launched with end-over-end monitoring, while later experiences were AB tested.\footnote{The template collection is continually expanding and so the maximum recall size varied for each experiment.}\ 

\begin{tabular}{ll}
\\
     1. & Reranking with External Image-Text Model (section \ref{sec:exp:clip}) \\
    2. & Null and Low recovery with Symbolic Multi-Modal Intents (section \ref{sec:exp:ckgnulllow})\\
    3. & Ranking with Domain-specific Image-Text Model (section \ref{sec:exp:adobeclip})\\
    4. & Recall with Sparse Image-Text Model (section \ref{sec:exp:rae})\\
    5. & Long Query Recall and Ranking with Multi-Modal Model (section \ref{sec:exp:mmckg})\\
    \\
\end{tabular}

\subsection{Reranking with External Image-Text Model}
\label{sec:exp:clip}

Our initial experiment used an external, English-only CLIP model in the rescore stage of the ranker.  To do this, we had to determine how many items to rescore.  This was largely governed by latency concerns since we wanted as many items as possible to use the CLIP multi-modal signal.  By running load tests, we determined that we could use the CLIP scores for the top 10K templates, where the initial ranking was determined by the existing ranker.

We also had to determine how to weight the CLIP scores in the rescoring.  Since the template ranker at this time was a non-ML, hand-tuned ranker, we determined this based on evaluations of a stratified query sample.  The extreme baseline was to use only the CLIP score for the reranking. This had two drawbacks: (1) The top results were not visually diverse enough, especially for broad queries like \textit{birthday card} or \textit{wedding invitation}; (2) there was not enough recent content to provide a sense of freshness and seasonality.  To determine a suitable weight, we  used a divide-and-conquer approach by starting with a 50\% balance between the first round ranker score and the CLIP score and then adjusting.  This quickly converged on a weighting of basically 2/3 for CLIP and 1/3 for the first round ranker score.  In AB testing, the click-through rate (CTR) and export rate improved with the CLIP-based reranking (Table \ref{tab:clip}).\footnote{All results are shown as percentage change. We are unable to show exact metrics.}\ There was no change in the null rate, as was to be expected.

\begin{table}[htb]
\caption{AB Test Results for CLIP Reranking. All results are statistically significant.}
\begin{tabular}{ll}
    Metric & Change \\
    \hline
    CTR & +7\%\\
    Export rate & +4\%\\
    \end{tabular}
    \label{tab:clip}
    \end{table}

\subsection{Null and Low Recovery with Symbolic Multi-Modal Intents}
\label{sec:exp:ckgnulllow}

Due to the broad range of user intents, the limited template collection, and the keyword based retrieval, users frequently landed on null and low result pages.  When the number of results are low ($<$5 results), the search engagement with the results drops significantly (up to 2--3x). To reduce the null and low result rate, we incorporated recovery mechanism using the symbolic CKG intents. The CKG intents for each template were indexed and the CKG intents for the query were calculated at query time. If there were $<$5 results, the CKG query intents were matched against the template intents. For example, the query \textit{hot yoga studio opening} has the intents \textit{yoga} and would match all templates with that intent. This resulted in major improvements in CTR and null rate (table \ref{tab:ckgnulllow}).

\begin{table}[htb]
\caption{AB Test Results for CKG Symbolic Intents Null and Low Recovery. All results are statistically significant.}
\begin{tabular}{ll}
    Metric & Change \\
    \hline
    Null rate & $-$69\%\\
    Null recovery CTR & +300\%\\
    Low rate & $-$59\%\\
    Low recovery CTR & +30\% \\
    Overall CTR & +7\%\\
\end{tabular}
\label{tab:ckgnulllow}
\end{table}


\subsection{Ranking with Domain-specific Image-Text Model}
\label{sec:exp:adobeclip}

The CLIP model (section \ref{sec:exp:clip}) only worked for English and was not optimized for Express templates and queries.  Replacing CLIP with AdobeCLIP  to rerank the top 10K Express templates was expected to be on par for English queries and improve the CTR for non-English queries. Because the recall and first-round ranker  constrained the result set, the core relevance, especially for head queries, was unlikely to change significantly, although the torso and tail queries, especially in non-English were expected be significantly different.
The move from CLIP to AdobeCLIP was part of a larger AB test which moved from an older search infrastructure to a newer one which, among other things, allowed for multiple embedding types. The goal of the AB test was to have no negative effects while moving to the new platform. This was borne out (table \ref{tab:adobeclip}).

\begin{table}[htb]
\caption{AB Test Results for Platform Move including CLIP-to-AdobeCLIP. Only the null rate change is statistically significant.}
\begin{tabular}{ll}
Metric & Change\\\hline
CTR & +0.0\%\\
Null rate & $-$7.7\%\\
Low rate & $-$5.0\%\\
\end{tabular}
\label{tab:adobeclip}
\end{table}



\subsection{Recall with Sparse Image-Text Model}
\label{sec:exp:rae}

None of the previous experiments leveraged the power of embeddings for augmenting the initial match set. The CLIP and AdobeCLIP models only affected the reranking of the search results. The null and low recovery with symbolic CKG multi-modal intents only affected null and low queries and leveraged symbolic intents.
Dense embeddings could not be used for the initial match set due to latency constraints. So, we experimented with using the AdobeCLIP sparse embeddings in the match set to augment the existing keyword matches.\footnote{We  considered using only AdobeCLIP sparse for the match set but rejected this since the model performed badly at identifying videos, which are popular queries and important to the business. AdobeCLIP only has the image embedding to match against and the images from Express video templates are extremely similar to those of still templates.}\  This required determining how many dimensions to match in the sparse embedding (section \ref{sec:clipAdobeCLIP}) in order to retrieve enough new relevant documents and not too many irrelevant ones.  As is well known in the literature \citep{ethayarajh2019contextualcontextualizedwordrepresentations,xiao2023lengthcurseblessingdocumentlevel} determining accurate thresholds on embeddings using cosine similarity or dot product is not feasible.\footnote{See \cite{rossietal2024} for a recent approach to determining relevance thresholds with embeddings.}\    The sparse embedding approach allowed us to require a minimum number of asset dimension matches.  Once the retrieval approach was determined, the ranking was updated to demote less relevant templates retrieved by the sparse embeddings (table \ref{tab:raerecall}).


\begin{table}[htb]
\caption{AB Test Results for  AdobeCLIP Recall and Ranking. All results are statistically significant.}
\begin{tabular}{lll}
Metric & \multicolumn{2}{c}{Change}\\ \hline
& English & Multilingual\\
CTR & +3\% & +3\% \\ 
Null Rate & $-$35\%& $-$14\% \\ 
Null and Low Recovery Rate & $-$54\% & $-$54\% \\
\end{tabular}
\label{tab:raerecall}
\end{table}


\subsection{Long Query Recall and Ranking with Multi-Modal Model}
\label{sec:exp:mmckg}

The above experiments improved relevancy for head queries, both in matching the  intent of the user query and in quality of the templates shown. In addition, the improved recall from CKG symbolic intents for null and low recovery (section \ref{sec:exp:ckgnulllow}) and the addition of sparse AdobeCLIP embeddings into the initial match set (section \ref{sec:exp:rae}) resulted in a broad set of related templates being shown when there are few exact matches.  However,  there are often few exact match templates for more specific user queries, i.e.\ for tail queries. 

To address this issue, we targeted longer ($>=$4 words) to use the CKG multi-modal embedding  (MM-CKG section \ref{sec:mmCKG}). 
The more specific intents of the longer queries work especially well with the domain-specific embeddings, allowing the recall and ranking to find the few templates that exactly match the user intent.  The ranking combined 1/3 the weight on MM-CKG and 2/3 the weight on AdobeCLIP. The hypothesis behind this was that AdobeCLIP captures the core relevance matching the query text to the image rendition, while MM-CKG captures the underlying intent of  the query and the template.  The optimal query length for this experience was determined empirically by manually judging a stratified sample of queries of different lengths, comparing production to the new experience. There was a clear demarcation between queries of $<$4 words and those of $>=$4 words.  Table \ref{tab:MMCKG} shows that for $<$4 words, both production and MM-CKG largely provide relevant results, i.e.\ for head queries both approaches work well.  However, for $>=$4 words the new MM-CKG results are significantly better than those in production.

\begin{table}[htb]
\caption{Relevancy results from human annotation when leveraging MM-CKG for recall and ranking.}
\includegraphics[width=3in]{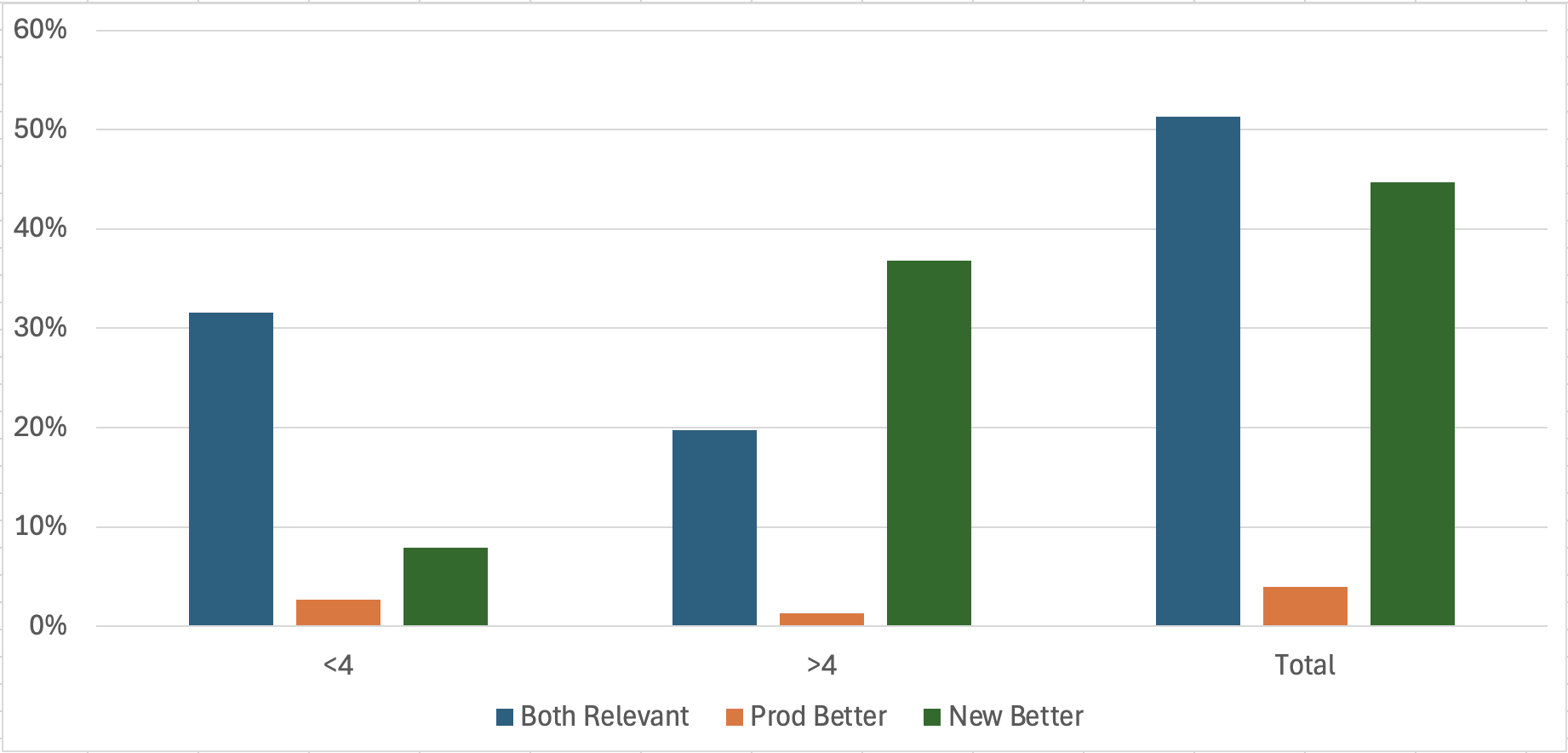}
\label{tab:MMCKG}
\end{table}

The MM-CKG embedding does not yet have a sparse version of the type available for AdobeCLIP (section \ref{sec:clipAdobeCLIP}). For this reason, it was not feasible from a latency perspective to use the MM-CKG matching and ranking approach used for all queries. However, since long queries are $<$10\% of the query traffic, we determined that the increased latency from calculating cosine similarity scores between the query and all of the templates was worth the improvements to relevance.

The AB test launched showcased statistically significant improvements of CTR and null rate on long queries and prompts, highlighting the usefulness of the hybrid system.

\begin{table}[htb]
\caption{AB Test Results for Long Prompt Understanding.}
\begin{tabular}{ll}
Metric & \multicolumn{1}{c}{Change}\\ \hline
CTR & +17\% \\ 
Null Rate & $-$46\% \\ 
\end{tabular}
\label{tab:longpromptab}
\end{table}

\section{Conclusion}
\label{sec:concl}

Multi-modal search experiences in industry applications traditionally depend on textual data in the index, thereby reducing the multi-modal search to a traditional keyword search. This provides a low latency experience since industry search engines are heavily optimized for keyword search. The advent of high quality multi-modal embeddings like CLIP  has provided radically new capabilities.  However, in an existing application, such as Adobe Express template search in this paper, the available multi-modal capabilities and the existing infrastructure including strict latency requirements, require a thoughtful, iterative approach to integrating new multi-modal technologies. This paper described five multi-modal experiments in Express template search, each of which built upon the others.  This has resulted in significantly lower null and low rates, while improving click-through rates.

\bibliography{mmsr}
\end{document}